\newcommand{\dint}{{\rm d}}

    \documentclass[%
  reprint,
  amsmath,amssymb,
  aps, nofootinbib
  ]{revtex4-1}
    \usepackage{amsmath}
    \usepackage{cleveref}
    \usepackage{amsfonts}
    \usepackage{graphicx}
    \usepackage{dcolumn}
    \usepackage{bm}

    \usepackage{titlesec}
  \usepackage{easyReview}
  \usepackage[bottom]{footmisc}

    \titlespacing{\subsubsection}{2pt}{\parskip}{-\parskip}
    
\begin{document}


    	\title{ Investigating the structure of gluon fluctuations in the proton with incoherent diffraction at HERA
	}

    	\author{Arjun Kumar}
    	\email{arjun.kumar@physics.iitd.ac.in}

    	\author{Tobias Toll}%
    	\email{tobiastoll@iitd.ac.in}
    	\affiliation{%
    		Department of physics, Indian Institute of Technology Delhi, India\\
    	}

    	\date{\today}

    	\begin{abstract}
    			Impact parameter dependent dipole models are ideal tools for investigating the spatial structure of the proton. We investigate the incoherent $ep$ cross section in exclusive $J/\psi$ photoproduction as measured by HERA, and find that as $|t|$ increases, the models need several levels of substructure of gluon density fluctuations in order to describe the measured data well. In lieu of a perturbative description, we add this substructure by hand. This substructure is modelled as hotspots within hotspots. This enables us to describe measurements for $|t|> 1$~GeV$^2$, which is necessary for describing any observable which integrate over the $t$-spectrum, such as the rapidity or $W_{\gamma p}$. We find that three levels of proton substructure are adequate for a good description of all available $ep$ data up to $|t|=30~$GeV$^2$. We note that the gluonic density fluctuation structure follows a scaling behaviour, such that the logarithms of the number of hotspots and their size fall on a line, effectively reducing the available parameter space of the model. Our findings systematically constrains and provides a benchmark for the development of a perturbative model of spatial gluon fluctuations in nucleons.

    	\end{abstract}

    	\maketitle


    	\section{\label{introduction}Introduction}
The dipole picture \cite{GolecBiernat:1998js, GolecBiernat:1999qd, Kowalski:2003hm, Kowalski:2006hc, Rezaeian:2012ji, Mantysaari:2018nng} has many advantages for the description of Deeply Inelastic Scattering (DIS) between electrons and protons. At small values of the Bjorken $x$ variable, it gives a clear interpretation of the underlying physics. In DIS, the virtual photon which probes the hadronic target splits up into a quark anti-quark dipole which subsequently interacts with the target via the strong force. The models based on the dipole picture are parametrisations which are fitted to inclusive HERA data. These models have a natural non-linear extension into a perturbative "saturation" regime in which strongly interacting matter becomes a Color Glass Condensate \cite{Gelis:2010nm,Iancu:2003xm}.
	
	The dipole-picture readily lends itself to an impact-parameter formulation, in which the transverse parton structure of the proton is described in coordinate space, while the longitudinal part is described with collinear parton density functions in momentum space. Thus far, the advantage of the impact-parameter dependent dipole models has been utilised in the study of heavy nuclear targets in preparation for the electron-ion collider (EIC) \cite{Accardi:2012qut,AbdulKhalek:2021gbh}. In diffractive events with heavy nuclei, the hadronic target stays intact during the interaction. However, in some cases, the target is excited in the collision and subsequently de-excites by emitting a photon, one or many nucleons, or by completely shattering into fragments. If the target remains intact the interaction is categorised as \emph{coherent} while if it gets exited the interaction is called \emph{incoherent}. The incoherent cross-section is proportional to quantum fluctuations in the amplitude of the interaction, while the coherent cross-section is proportional to its expectation value. The impact parameter is a Fourier conjugate to $\Delta=\sqrt{-t}$, where $t$ is the Mandelstam variable, which means that at large $|t|$ one may resolve transverse gluon fluctuations at smaller length scales. For $e$A collision, the incoherent cross-section has been implemented in the dipole model through event-by-event variations in the configurations of constituent nucleon position in the nucleus \cite{Toll:2012mb}, taking full advantage of the transverse coordinate description of the dipole models. The nucleons are assumed to fall on a Woods-Saxon distribution and the gluons inside the nucleons are assumed to follow Gaussian distributions, which is the result of independently fluctuating partons. This approach has been seen to underestimate the incoherent cross-section in confrontations with measurements of ultra-peripheral collisions at the LHC \cite{Sambasivam:2019gdd}. 

	In 2016, M\"antysaari and Schenke extended the event-by-event gluon fluctuations by taking them inside the proton \cite{Mantysaari:2017dwh, Mantysaari:2016ykx,Mantysaari:2016jaz}. They found that a good description of proton dissociation measurements could be found if the gluons in the proton are not isotropically distributed but are located in the vicinity of three "hotspots", each with a Gaussian shape in the transverse plane. This can be understood as the evolution of a few gluons with large momentum fractions $x$ which evolve by splitting into many gluons of smaller $x$, and that this evolution retains spatial correlations with the original gluons. The picture which emerges is one where at large length-scales the transverse gluon density of the proton is described by a single Gaussian, but when probed at larger $|t|$ it contains three smaller hotspots. They found a good agreement with HERA data using the impact-parameter dependent dipole model. Following their work, different variants of hotspot models were studied \cite{Traini:2018hxd,Cepila:2017nef,Cepila:2016uku,Cepila:2018zky,Bendova:2018bbb,Demirci:2022wuy} explaining the incoherent $ep$ data. It is expected that these subnucleon fluctuations will play an important role for $e$A collisions at the EIC as well, for larger values of $|t|$. For a recent review on subnuclear fluctuations, see \cite{Mantysaari:2020axf}. These subnucleon fluctuations have also been studied in context of hydrodynamic simulations of p+Pb collisions where significantly larger values of the azimuthal flow coefficients \textit{v}$_{n}$ were obtained compared to those of protons without fluctuating substructure \cite{Mantysaari:2017cni, Schenke:2021mxx}. The  hotspot profile of the proton has also been used to investigate the correlations and hollowness in proton-proton interactions at high energies \cite{Albacete:2016pmp,Albacete:2016gxu,Albacete:2017joc,Albacete:2018rzf}. 
	
	The shortcoming thus far of the hotspot model is that it is a non-perturbative model and thus is valid in the low momentum transfer region, where $|t|\lesssim 1~$GeV$^2$. This is a problem for several reasons, e.g. when calculating rapidity- or invariant mass spectra, $t$ is integrated over, and thus the total and incoherent cross-sections are significantly under-estimated for these observables. Also, for the saturated dipole model, the hotspot model fails to reproduce the coherent cross section, as well as the very small $|t|$ region of the incoherent spectrum (M\"antysaari and Schenke describe the small $|t|$ region by combining the dipole model with a glasma model \cite{Mantysaari:2016ykx,Mantysaari:2016jaz}). Furthermore, since HERA experiments have measured the $ep$ $t$-spectrum up to $|t|=30$~GeV$^2$, a description of this physics is desirable. In this paper, a refined hotspot model is proposed which explains the HERA incoherent data in both low and high $|t|$ regions and investigate the fluctuations in both the saturated and non-saturated models. In lieu of a perturbative description of the spatial gluon structure of the proton, we investigate what such a theoretical model should look like, as we expect that the hotspot themselves will exhibit substructure which can be probed at larger $|t|$. We show that the large $|t|$ data can be well described in a model with hotspots within hotspots, where the three hotspots each have a substructure of even smaller spatial regions of gluon density fluctuations. If we further add more substructure within the smaller hotspots we find a good description of all available HERA incoherent data with $|t|<30~$GeV$^2$. We find that the resulting structure of these quantum fluctuations exhibit self-similarities. This can be used to constrain the behaviour of a future perturbative model of gluon substructure and to make prediction for future measurements at the EIC or a future LHeC experiment \cite{Agostini:2020fmq}. We also suggest a new proton thickness profile which retains the coherent cross section in the saturated dipole model described below, and which improve the small-$|t|$ description of the incoherent data by introducing one extra parameter related to the level of gluon-gluon correlations in the proton. We primarily do this with two versions of the impact-parameter dependent dipole model, which we supplement as the study progresses: the saturated bSat model, and its linear version, the bNonSat model. Both models have been shown to describe existing $ep$ measurements well, while recent UPC measurements at the LHC clearly prefer the bSat model \cite{Sambasivam:2019gdd}.
	
	The paper is organised as follows. In the next session we give a brief description of the dipole models we use in this paper. In section \ref{fluctuations} we describe the hotspot model and our addition of extra substructure to it. In section \ref{results} we present the results from the models, and compare it to $ep$ measurements. Finally in section \ref{conclusion} we discuss the results and how to take this project further. 
	
    	\section{\label{dipole_picture} The Color Dipole Models}

\begin{figure}
	\centering
	\includegraphics[width=0.65\linewidth]{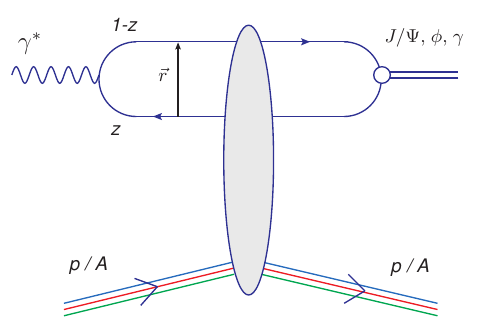}
	\caption{Exclusive vector meson production in the dipole picture of DIS. See description in text. }
	\label{dipole}
\end{figure}

The scattering amplitude for the diffractive vector meson production is a convolution of three subprocess, as depicted in Fig.~\ref{dipole}. First, the virtual photon splits into a quark anti-quark dipole, then the dipole interacts with the proton via one or many colourless two-gluon exchanges, after which it recombines into a vector meson. The amplitude is given by:\footnote{Note that we do not use the corrected phase-factor $(\frac12-z)\vec r\cdot\vec \Delta$ as was found in \cite{Hatta:2017cte, Mantysaari:2020lhf}. We have checked that our results are not affected by this change beyond the precision of the models.}
 \begin{eqnarray}	\label{eq:amplitude}
    	\mathcal{A}_{T,L}^{\gamma^* p \rightarrow J/\Psi p} (x_{\mathbb{P}},Q^2,{\bf \Delta})
	&=&i\int \dint^2 {\bf r}\int \dint^2{\bf b}\int \frac{\dint z}{4 \pi}  \nonumber \\
	&\times& (\Psi^*\Psi_V)_{T,L}(Q^2,{\bf r},z) \\  
	&\times& e^{-i[\bf{b}-(1-z){\bf r}].{\bf \Delta}} \frac{\dint\sigma _{q\bar{q}}}{\dint^2{\bf b}}({\bf b},{\bf r},x_\mathbb{P})\nonumber
\end{eqnarray}
 where $T$ and $L$ represent the transverse and longitudinal polarisation of the virtual photon, \textbf{r} is the transverse size of the dipole, \textbf{b} is the impact parameter of the dipole relative to the proton, $z$ is the light-cone momentum fraction of the photon taken by the quark and ${\bf \Delta} \equiv \sqrt{-t} $ is the transverse part of the four-momentum difference of the outgoing and the incoming proton. $(\Psi^* \Psi_V)$ denotes the wave-function overlap between the virtual photon and the produced vector meson. It should be noted that this amplitude is a Fourier transform from transverse coordinate of the quark in the dipole, to the transverse momentum transfer of the proton. Hence, the resulting cross section contains information on the spatial structure of the proton. The dipole cross section $\frac{d\sigma _{q\bar{q}}}{d^2\textbf{b}}(\textbf{b},\textbf{r},x_\mathbb{P})$ describes the strong interaction. The vector-meson wave function are usally modelled and we use boosted Gaussian wavefunction for $J/\psi$  with the parameter values from \cite{Mantysaari:2018nng}. A recent update on different vector-meson wavefunctions can be found in \cite{Lappi:2020ufv}.
    The elastic diffractive cross section for a spherical proton (without geometrical fluctuations), is given by:
\begin{equation}
    	\frac{d \sigma^{\gamma^* p \rightarrow J/\Psi p}}{dt} = \frac{1}{16 \pi} \big| \mathcal{A}^{\gamma^* p \rightarrow J/\Psi p} |^2
\end{equation}
 When fluctuations are included in the wave function of the proton, e.g. due to different number of interacting constituents, or variation of transverse positions of constituents, we employ the Good-Walker formalism \cite{Good:1960ba}. The coherent cross section probes the first moment of the amplitude which corresponds to its average, while the total cross section probes the second moment. The incoherent cross section thus probes the difference between the second moment and first moment squared, which for Gaussian distributions is its variance. Thus, for an event-by-event variation $\Omega$, we have:
    \begin{eqnarray}
    	\frac{{\rm d} \sigma_{\rm coherent}}{{\rm d}t} &=& \frac{1}{16 \pi} \big| \left<\mathcal{A}(x_{\mathbb{P}},Q^2,\textbf{$\Delta$})\right>_\Omega\big|^2 \nonumber \\
    	\frac{{\rm d} \sigma_{\rm incoherent}}{{\rm d}t} &=& \frac{1}{16 \pi}\bigg(\big< \big| \mathcal{A}(x_{\mathbb{P}},Q^2,\textbf{$\Delta$})\big|^2\big>_\Omega \\\nonumber&~& ~~- \big| \big<\mathcal{A}(x_{\mathbb{P}},Q^2,\textbf{$\Delta$})\big>_\Omega\big|^2\bigg)
    	\end{eqnarray}
Fluctuations should therefore be introduced in a way which retains their average distribution and hence the coherent cross section, and only contributes to the incoherent cross section.

We consider two versions of the dipole cross-section.
The bSat model dipole cross section is given by \cite{Bartels:2002cj}:
 \begin{eqnarray}
    	\frac{\dint\sigma _{q\bar{q}}}{\dint^2\textbf{b}}(\textbf{b},\textbf{r},x_\mathbb{P})=
	2\big[1-\text{exp}\big(-F(x_\mathbb{P} ,\textbf{r}^2)T_p(\textbf{b})\big)\big]
\end{eqnarray}
with
\begin{eqnarray}
    	F(x_\mathbb{P} ,\textbf{r}^2)=\frac{\pi^2}{2N_C} \textbf{r}^2 \alpha_s(\mu^2) x_\mathbb{P} g(x_\mathbb{P},\mu^2),
\end{eqnarray}
Due to the exponential form, this formulation saturates the cross-section as the gluon density $xg(x, \mu^2)$ becomes large as well as for large dipole sizes $r$. 
The scale at which the strong coupling $\alpha_s$ and gluon density is evaluated at is $\mu^2 = \mu_0^2 +\frac{C}{r^2}$ and the gluon density at the initial scale $\mu_0$ is parametrised as:
\begin{eqnarray*}
   x g(x,\mu_0^2)= A_g x^{-\lambda_g}(1-x)^{6}
\end{eqnarray*}
where the parameters $A_g$, $\lambda_g$, $C$, and $m_f$ are determined through fits to inclusive DIS reduced cross section measurements. We use the fit results from \cite{Sambasivam:2019gdd}. The transverse profile of the proton is usually assumed to be Gaussian:
\begin{eqnarray}
    	T_p(\textbf{b}) = \frac{1}{2 \pi B_G}\exp\bigg(-\frac{\textbf{b}^2}{2B_G}\bigg)
\label{eq:profile}
\end{eqnarray}
and the parameter $B_G$ is constrained through a fit to the $t$-dependence of the exclusive J/$\psi$ production at HERA \cite{Kowalski:2006hc}, and is found to be $B_G=4\pm 0.4~$GeV$^{-2}$. 

The linearised dipole cross section is called the bNonSat model:
\begin{equation}
    	\frac{\dint\sigma _{q\bar{q}}}{\dint^2\textbf{b}}(\textbf{b},\textbf{r},x_\mathbb{P})=\frac{\pi^2}{N_C}\textbf{r}^2\alpha_s(\mu^2) x_\mathbb{P} g(x_\mathbb{P},\mu^2)  T_p(\textbf{b})
\end{equation}
which does not saturate for large gluon densities and large dipoles. This dipole cross-section corresponds to a single two-gluon exchange. 
At present both models describe measured HERA $F_2$ and exclusive data well \cite{Mantysaari:2018nng,Rezaeian:2012ji}, while studies of ultra-peripheral collisions of lead nuclei at the LHC show a clear preference for the bSat model \cite{Sambasivam:2019gdd}.
\begin{figure}
	\centering
	\includegraphics[width=0.8\linewidth]{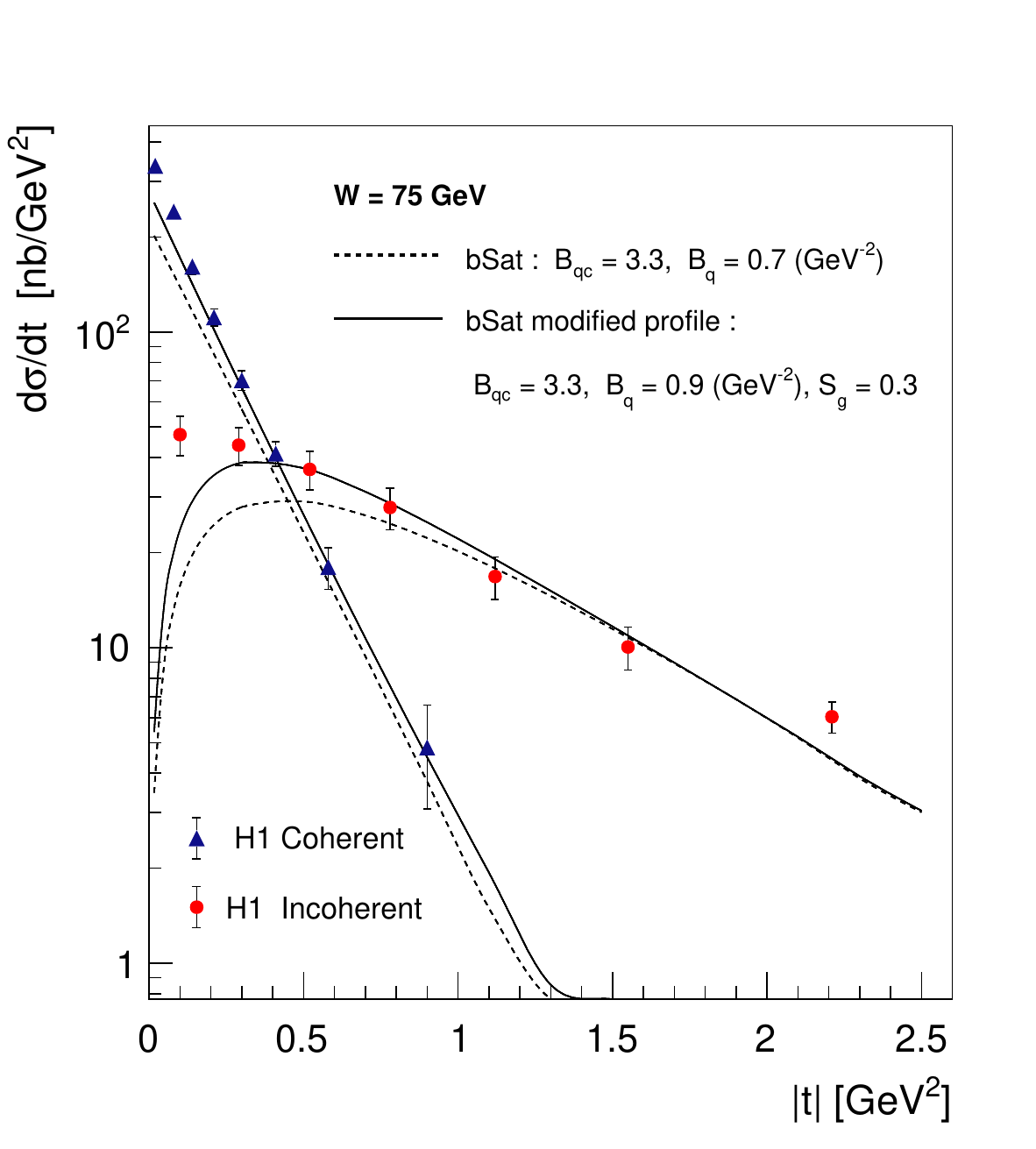}
	\caption{The $|t|$ dependence of $J/\Psi$ photoproduction in the bSat hotspot model with and without modified profile function, compared to measurements from H1 \cite{Alexa:2013xxa}.}
	\label{mod_bSat}
\end{figure}
The exclusive cross section receives large corrections (discussed in detail in \cite{Kowalski:2006hc, Mantysaari:2016ykx}). Firstly, the scattering amplitude in eq.~\eqref{eq:amplitude} is approximated to be purely imaginary. However, the real part of the amplitude is taken into account  by multiplying the cross section by a factor of (1+$\beta^2$) with $\beta = \tan\big(\lambda \pi/{2})$, and $\lambda = \partial \log (\mathcal{A}_{T,L}^{\gamma^*p\rightarrow Vp})/\partial \log(1/x)$. Secondly, to take into account that the two gluons may have different momentum fractions, a skewedness correction to the amplitude is introduced \cite{Shuvaev:1999ce}, by a factor $R_g(\lambda)= 2^{2 \lambda +3}/\sqrt{\pi} \cdot\Gamma (\lambda_g + 5/2)/\Gamma (\lambda_g+4)$ with $\lambda_g= \partial \log (xg(x))/\partial \log(1/x)$.
In our model, we calculate both these corrections using a spherical proton. They have significant contribution at low $|t|$ (around 40-60\%) while at large momentum transfers their contribution dwindle.
 \section{ \label{fluctuations} Geometric Fluctuations in the Proton Wave Function}
There may be many different sources of quantum fluctuations in the proton wave-function, e.g event-by-event variations in the transverse positions of constituent partons or fluctuations in the number of constituents itself. In this section we explore various geometrical fluctuations in detail.

	\begin{figure}
	\centering
	\includegraphics[width=0.85\linewidth]{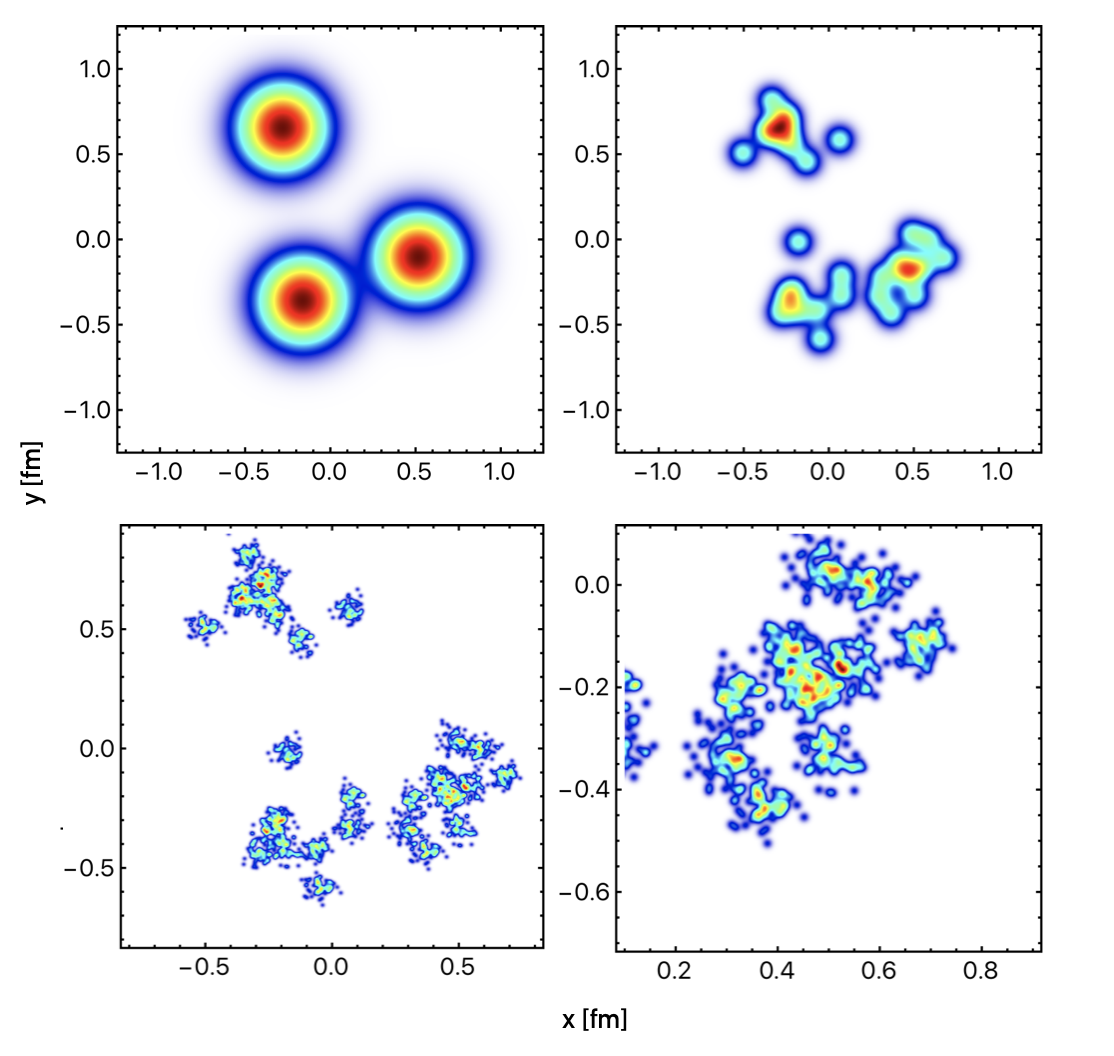}
	\caption{Transverse profile of the proton in an event in the refined hotspot model with the three first panels adding one level of substructure. The fourth panel is an enhancement of the third panel. The parameters are from the bSat model in table \ref{table}.}
	\label{profile}
\end{figure}    
	
The transverse profile of the proton is assumed to be made of hotspots which can be implemented by changing the proton profile in eq.\eqref{eq:profile} thus \cite{Mantysaari:2016ykx}:
\begin{eqnarray}
    	T_p(\textbf{b}) \rightarrow \frac{1}{N_q}\sum_{i=1}^{N_q}T_q(\textbf{b-b$_i$}),
\end{eqnarray}
with

\begin{equation}
    	T_{q}(\textbf{b}) = \frac{1}{2 \pi B_{q}}\exp\big[-\frac{\textbf{b}^2}{2B_{q}}\big]
\label{eq:hsprofile}
\end{equation}
Here $N_q$ = 3 and $\textbf{b}_i$ are the locations of hotspots sampled from a Gaussian width $B_{qc}$ and $B_q$ is the width of the hotspots.  $B_{qc}$, and $B_q$ control the amount of fluctuations in the proton geometry at low momentum transfer and are constrained by the coherent and incoherent data. 

In the bNonSat model, the average of the amplitude over configurations is directly proportional to the average of the profile function i.e
\begin{equation}
    \nonumber \big<\mathcal{A}\big>_\Omega \propto \big<T_P(b)\big>_\Omega
 \end{equation}
while this is \emph{not} the case for the bSat model which has a non-linear dependence on the proton profile. As a result, the coherent cross-section is reduced as shown in fig \ref{mod_bSat}. One can solve this by finding a new profile function which retains the spherical average of the amplitudes in the hotspot model. Here, the profile of the hotspots in the bSat model is modified as follow:
\begin{equation}
    	T_q(\textbf{b}) = \frac{1}{2 \pi B_q}\frac{1}{\big(\exp\big[\frac{\textbf{b}^2}{2B_q}\big]-S_g\big)}.
\end{equation}
If the parameter $S_g=0$ the profile remains a Gaussian, while with $S_g>0$ it becomes more peaked at the centre. We may interpret this parameter as being related to probed correlations between gluons within the hotspots. We may also note that this profile is no longer normalised to unity for $S_g\neq 0$. For the bSat model one therefore needs to choose between retaining the coherent cross section or the profile normalisation when adding substructure to the proton. Due to the non-linearity of the bSat model we cannot make a straightforward probabilistic interpretation to the thickness function, as is the case in bNonSat. Therefore, we prefer to preserve the coherent cross section while adding proton substructure. As will be shown, both this modified bSat model and the bNonSat model behave very similarly, which further validates this approach. 

\begin{figure}
	\centering
    		\includegraphics[width=1.05\linewidth]{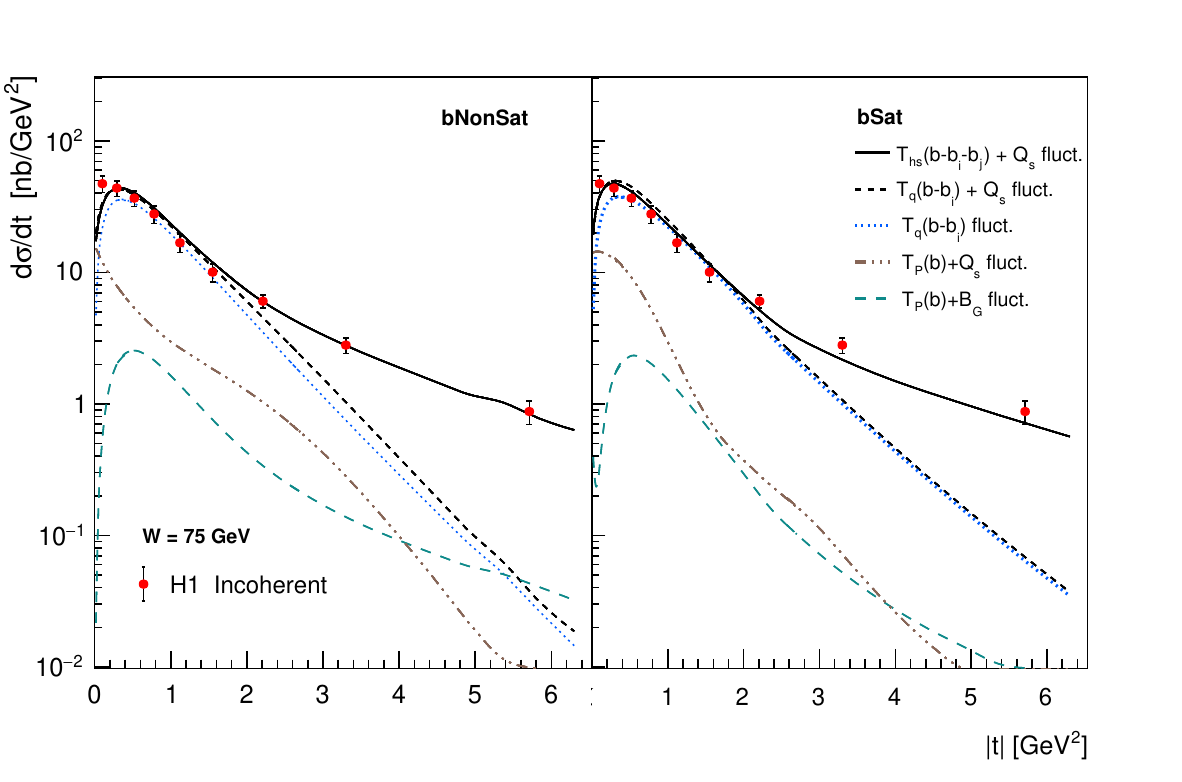}
    	\caption{The $|t|$ dependence of incoherent $J/\Psi$ photoproduction in refined hotspot model with separate contributions from different sources of fluctuations as described in the text, in the bNonSat model (left) and bSat model with modified profile (right).}
    			\label{master}
 \end{figure}

As the hotspot model is a non-perturbative description of the proton's structure it is not reliable for $|t|\gtrsim1$~GeV$^2$. However, as can be seen in the right panel of Fig. \ref{master}, it describes the data well for $|t|\lesssim 2.5$~GeV$^2$. As the $|t|$ distribution at leading twist is a Fourier transform of the fluctuating structure of the proton, this indicates that no further structure than the original hotspot contributes significantly to the incoherent cross section. However, for larger values of $|t|$, the measured $t$-spectrum exhibits two distinct slopes. As we increase $|t|$, we expect to resolve smaller scale gluon fluctuations in the transverse plane. This hints toward the presence of substructure of the hotspots (we will also explore non-Gaussian shapes as an alternative explanation below). As we presently do not have a perturbative method for adding further substructure, we do so by hand. 

We thus introduce smaller hotspots within the original three hotpsots, which  modifies the hotspot density in eq.~\eqref{eq:hsprofile} as follows:
  \begin{table*}
	\begin{tabular}{|c|c|c|c|c|c|c|c|c|c|}
		\hline
		\textbf{Model} &\textbf{ B$_{\textbf{qc}}$} &  \textbf{ B$_{\textbf{q}}$} & \textbf{ N$_{\textbf{q}}$}&  \textbf{ B$_{\textbf{hs}}$} & \textbf{ N$_{\textbf{hs}}$} & \textbf{ B$_{\textbf{hhs}}$}& \textbf{ N$_{\textbf{hhs}}$} &  \textbf{ S$_{\textbf{g}}$}&  $\sigma$ \\
		\hline
		bNonSat hotspot& 3.2 &  0.9&3& -- & -- & -- &-- &--&0.4 \\
		\hline
		bSat  hotspot	& 3.3 &  0.7&3& -- &--  & -- &--&--& 0.5 \\
		\hline
		modified bSat  hotspot	& 3.3 & 0.9 &3& --  & -- &-- &-- & 0.3 &0.4 \\
		\hline
		{ bNonSat refined hotspot}	&  { 3.2} & { 1.15} &3& { 0.05} &{ 10} &--&-- & --  & { 0.4} \\
		\hline
		{ bSat refined hotspot}	& { 3.3} & { 1.08} &3& { 0.09}&{ 10} &--&--  & { 0.4}  &{ 0.5} \\
		\hline
		{ bNonSat further refined hotspot}	&  { 3.2} & { 1.15} & { 3}&{ 0.05}  &{ 10} &{ 0.0006}&{ 65} & -- & { 0.4} \\
		\hline
		{ bSat further refined hotspot}	& { 3.3} & { 1.08}&{ 3} & { 0.09}  &{ 10} &{ 0.0006}&{ 60}& { 0.4} &{ 0.5} \\
		\hline
	\end{tabular}
	\caption{Parameter values in the different dipole models described in the text. All the $B_i$ parameters are in GeV$^{-2}$.}
	\label{table}
\end{table*}
\begin{equation}
    	T_q(\textbf{b}) \rightarrow \frac{1}{N_{hs}}\sum_{j=1}^{N_{hs}}T_{hs}(\textbf{b-b$_j$})
\end{equation}
where $T_{hs}(\textbf{b})$ is given by :
\begin{equation}
    	T_{hs}(\textbf{b}) = \frac{1}{2 \pi B_{hs}}\exp\big[-\frac{\textbf{b}^2}{2B_{hs}}\big]
\end{equation}

The measured incoherent $t$-spectrum changes slope again for $|t|\simeq 12.5~$GeV$^2$. In order to take this into account we introduce yet another level of substructure to the gluon fluctuations, with density profile:
\begin{eqnarray}
    	T_{hhs}(\textbf{b}) = \frac{1}{2 \pi B_{hhs}}\exp\big[-\frac{\textbf{b}^2}{2B_{hhs}}\big]
\end{eqnarray}
 The proton profile function in one event in this further refined hotspot model is then given by:
\begin{eqnarray}
	T_{P}(\textbf{b}) &=&
	 \frac{1}{2 \pi N_q N_{hs}N_{hhs} B_{hhs}} 
	 \sum_{i}^{N_q}\sum_{j}^{N_{hs}}\sum_{k}^{N_{hhs}}
	e^{-\frac{(\textbf{b}-\textbf{b}_i-\textbf{b}_j-\textbf{b}_k)^2}{2B_{hhs}}}\nonumber
\end{eqnarray}
Here, \textbf{b}$_i$, \textbf{b}$_j$ and \textbf{b}$_k$ determine the transverse positions of the larger, smaller and smallest hotspots respectively, which fluctuates event-by-event. We have introduced four new parameters, the number of smaller hotspots in each level of substructure, $N_{hs}$ and $N_{hhs}$, and the width of these hotspots $B_{hs}$ an $B_{hhs}$.  We find a reasonable convergence when averaging over one thousand hotspot configurations. The resulting proton profile is illustrated in Fig.~\ref{profile} for one, two, and three levels of substructures. It may appear as we are introducing many new parameters to describe a few data-points, but we will show below that the $B$ and $N$ parameters are highly correlated, and that the proton's substructure exhibit a scaling behaviour.


\begin{figure}
	\centering
	\includegraphics[width=1.02\linewidth]{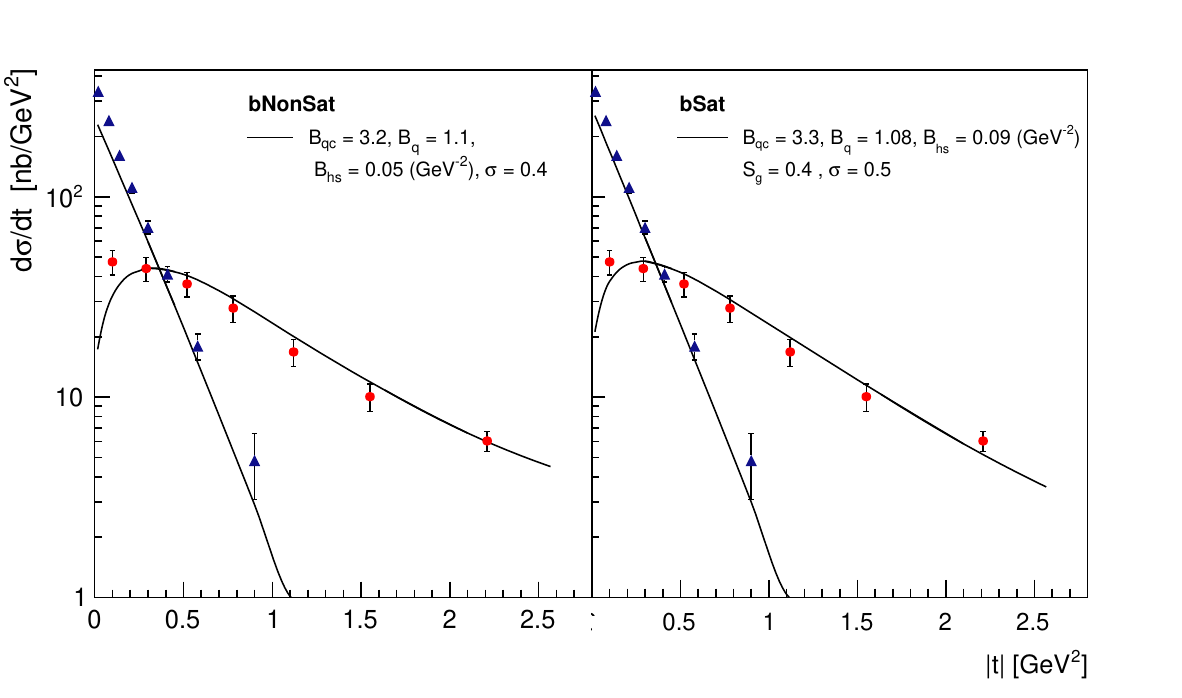}\hskip01cm
	\caption{The $|t|$ dependence of the coherent and incoherent J/$\Psi$ photoproduction in the saturated and non-saturated version of the refined hotspot model. }
	\label{coherent}
\end{figure}
An additional source of fluctuations is due to the fluctuations in the saturation scale. Following \cite{Mantysaari:2016ykx}, the saturation scale fluctuations are incorporated by letting the saturation scale of larger hotspots fluctuate independently. These are the fluctuations of the normalisation of the hotspots. These fluctuations are motivated by the experimentally observed multiplicity distributions and the rapidity correlations in p+p collisions which require the saturation scale to fluctuate according to the following distribution \cite{Bzdak:2016aii,McLerran:2015qxa}:
\begin{align}
    	P(\text{ln } Q_s^2/\big<Q_s^2\big>)=\frac{1}{\sqrt{2 \pi \sigma^2}} \text{exp}\bigg[-\frac{\text{ln}^2 Q_s^2/\big<Q_s^2\big> }{2  \sigma^2}\bigg]
\end{align}
Since the saturation scale is $Q_S^2(x, b)\equiv 2/r^2_S$, where $r_S$ solves the equation $1/2=F(x, r_S^2)T_p(b)$, in the dipole amplitude, we can implement these fluctuations by changing the normalisation of the profile function. We also have to divide the profile function by the expectation value $\big<E\big>=\exp(\sigma^2/2)$ of the log-normal distribution due to the increased average of the saturation scale due to the sampling. This introduces another variable $\sigma$ which is restricted by the data. These fluctuations play an important role in the low $|t|$ region while for the large momentum transfer which is the focus of this work, it is the geometrical fluctuations of smaller hotspots that dominate. 

    	Another possible contribution to the small $|t|$ spectrum could be proton size fluctuations, which can be described by the event-by-event fluctuations of the proton size parameter $B_G$. For completeness, we implement these fluctuations by sampling the mean value of $B_G$ as $4$~GeV$^{-2}$ with a variance of $1.3 $~GeV$^{-2}$ which controls the amount of fluctuations. The variance chosen here is rather large, to show the limit of the contribution of these kinds of fluctuations. These are large length-scale fluctuations and are only expected to contribute to the cross-section at small $|t|$, but as will be seen, even with this large variance this contribution is negligible. 

\begin{figure}
	\centering
	\includegraphics[width=1.1\linewidth]{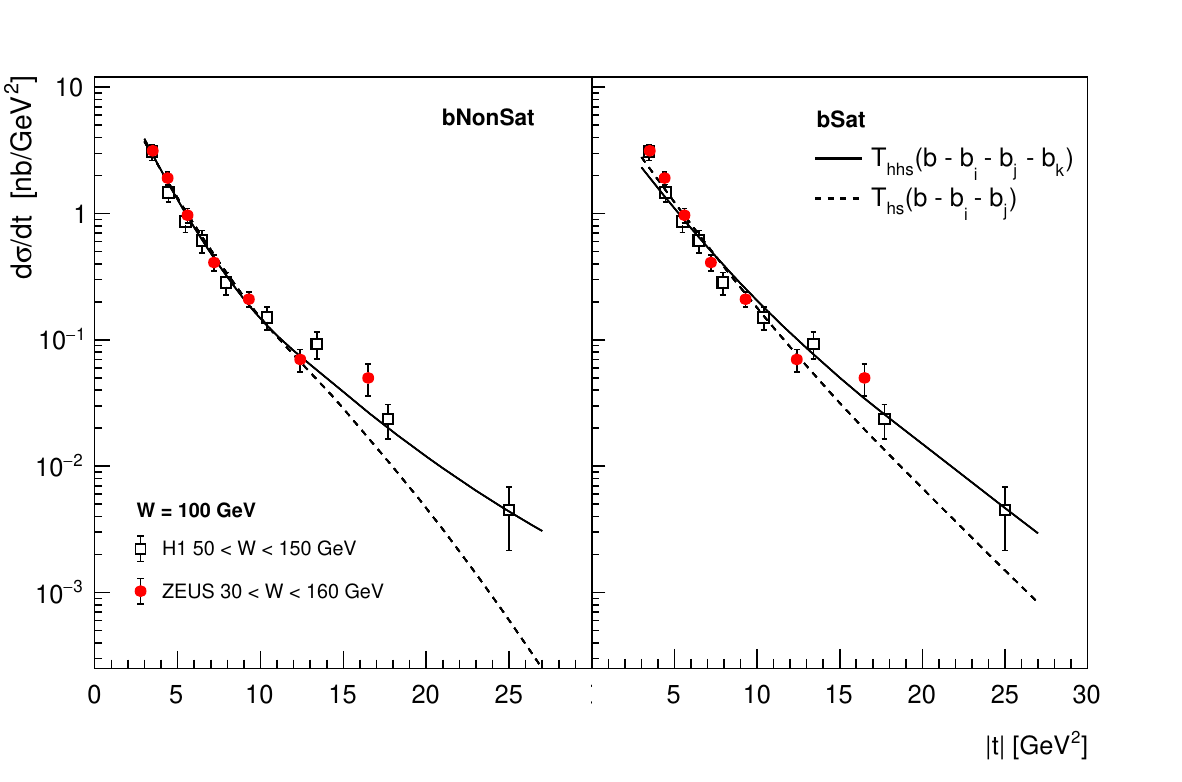}
	\caption{Contributions from two and three levels of hotspot substructure to the incoherent $|t|$ spectrum with the bNonSat model (left) and the bSat model (right). }
	\label{fig:EICprediction}
\end{figure}

\section{Results}\label{results}

In table \ref{table}, we show the values we have found for all the parameters described above. The second line is identical with earlier results \cite{Mantysaari:2016jaz} and the bNonsat model prefers slightly larger hostspot size 
as compared with the bSat model. 

Fig.~\ref{profile} depicts the snapshot of the resulting transverse profile of an event, where each panel adds one level of substructure, using the parameters for the bSat model. The last panel is a zoomed in version of the previous one. 

In Fig.~\ref{mod_bSat}, we find not only that the bSat model with the modified profile describes the coherent data well, but also that the description of the incoherent data for small $|t|$ improves. This gives us confidence in our choice of modified proton profile. For the rest of the paper we will refer to this modified bSat model as bSat.

\begin{figure*}
	\centering
	\includegraphics[width=0.4\linewidth]{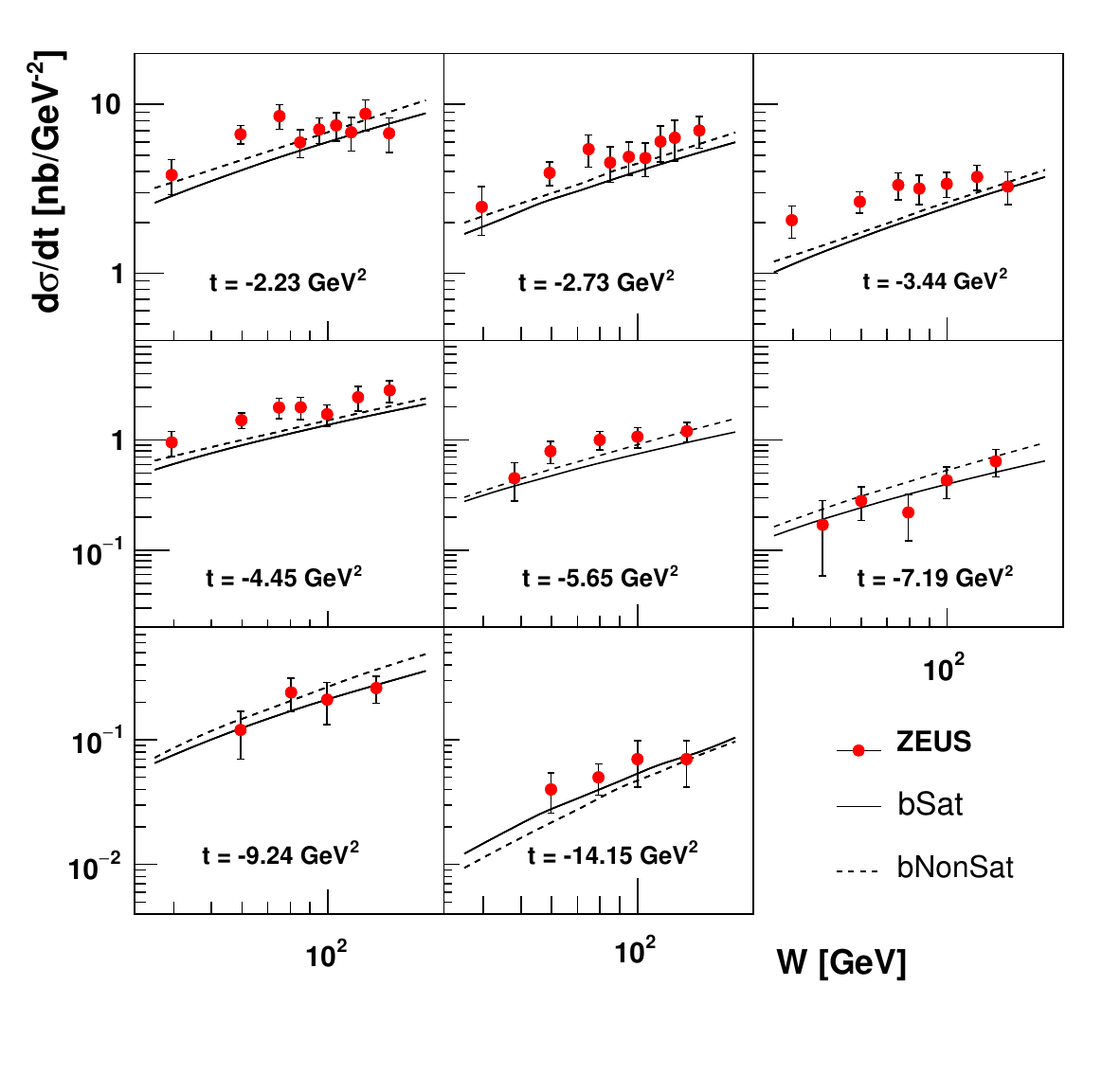}\hskip01cm
	\includegraphics[width=0.335\linewidth]{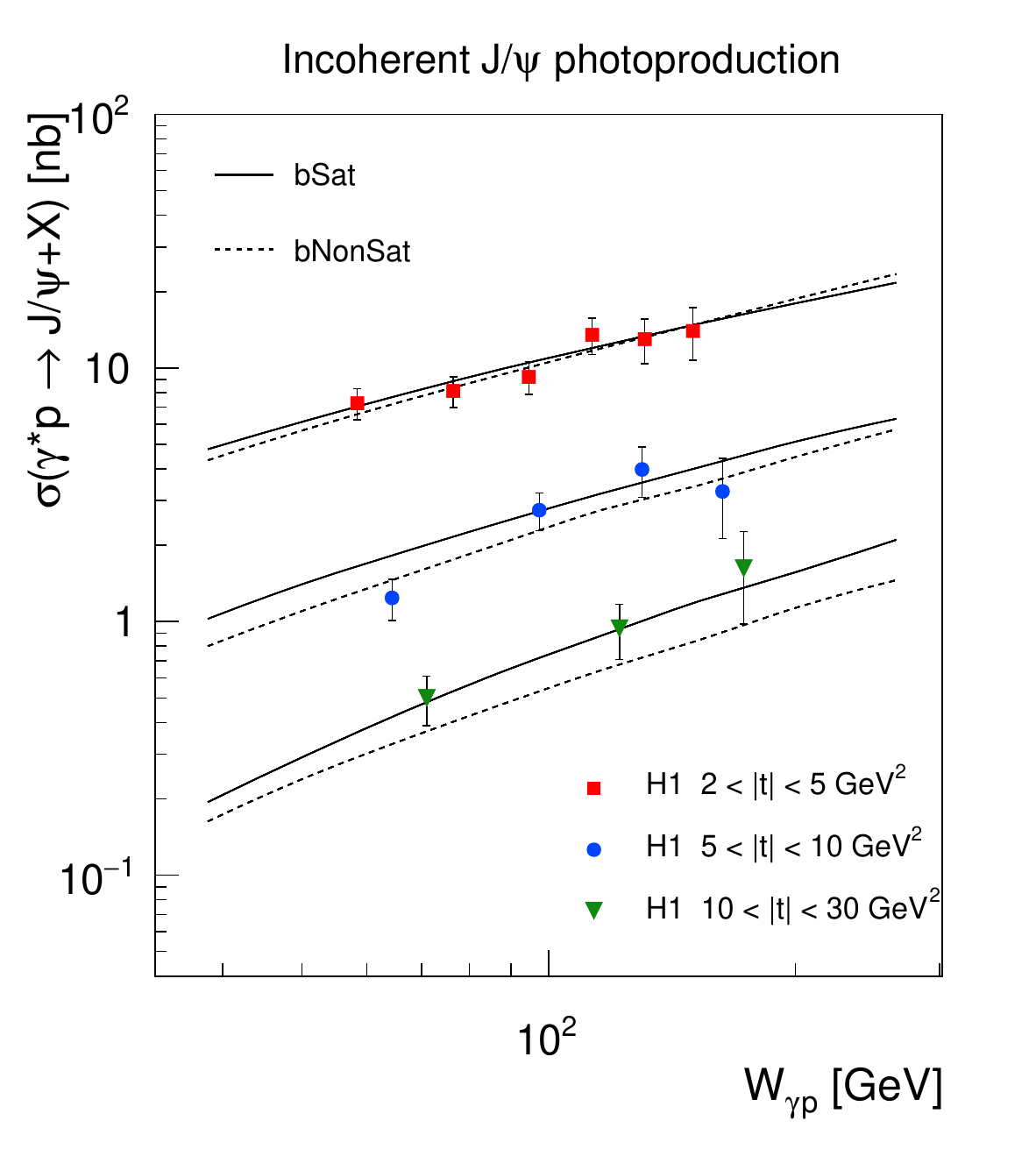}
	\caption{Energy dependence of the incoherent J/$\Psi$ production in the saturated and non-saturated version of the refined hostpot model. The ZEUS data \cite{Chekanov:2002rm,Chekanov:2009ab} for $|t|$ and energy dependence is available in the kinematic range $ 30< W <160~$ GeV  and $2<|t|<20~$GeV$^2$ while the H1 \cite{Aktas:2003zi} has measured the differential cross section $d\sigma(\gamma p \rightarrow J/\Psi Y)/dt$ in the range $2<|t|<30~$GeV$^2$ and the total cross section as a function of the photon-proton centre-of-mass energy in the range $50< W <200~$ GeV.}
	\label{energy_dep}
\end{figure*}
In Fig.~\ref{master} we show the incoherent cross section for the models considered against the photoproduction measurement of $J/\Psi$-mesons at $W=75$ GeV from H1 \cite{Alexa:2013xxa}, for both the saturated and non-saturated dipole models. Proton size fluctuations are seen to give an insignificant contribution to the incoherent cross section, and thus we can safely ignore this geometrical fluctuation henceforth. We see that the larger hotspot model agrees well with the data in the low momentum transfer region $|t|<2.5~$GeV$^2$, but underestimates the data points at very low $|t|$ values. The smallest $|t|$ region with $|t|<0.3~$GeV$^2$ is improved by the inclusion of saturation scale fluctuations. We also show the separate contribution from the saturation scale fluctuations. As discussed above, the bSat model contains one more parameter in its modified proton shape function, which improves the small $|t|$ description of the incoherent spectrum. For $|t| > 2.5~$GeV$^2$ the data is only well described by the refined hotspot model which contains ten hotspots within each of the original three hotspots. As can be seen in the figure, including fluctuations of the saturation scale, as well as larger and smaller hotspots gives a good description of the measured data for the entire spectrum of $|t|<7~$GeV$^2$.

In Fig. \ref{coherent}, we show the measured coherent cross-section \cite{Alexa:2013xxa}. Here, the coherent cross-section is well described by both models. Note that the hotspots in the bSat model has a modified profile. As desired, we see that the smaller hotspots do not affect the coherent cross-section.

In Fig.~\ref{fig:EICprediction} we show the resulting $t$-spectrum for very large $|t|$ for both the saturated and non-saturated versions of the dipole model. Here we see that two levels of substructure is not enough to describe the data when $|t|\gtrsim 12.5$~GeV$^2$. The description of the data improves significantly by the inclusion of one extra level of substructure. 

In Fig.~\ref{energy_dep} we study how the J/$\Psi$ photoproduction cross section varies with the photon-proton centre of mass energy $W_{\gamma p}$ in different bins of $t$, as compared with with the H1 \cite{Aktas:2003zi} and ZEUS \cite{Chekanov:2009ab} measurements. The cross section increases with increasing $W_{\gamma p}$ for all ranges of $|t|$-values and the model predictions are in good agreement with the measurements for both the bSat and bNonSat models.

 \begin{figure*}
 	\centering
 	\includegraphics[width=0.273\linewidth]{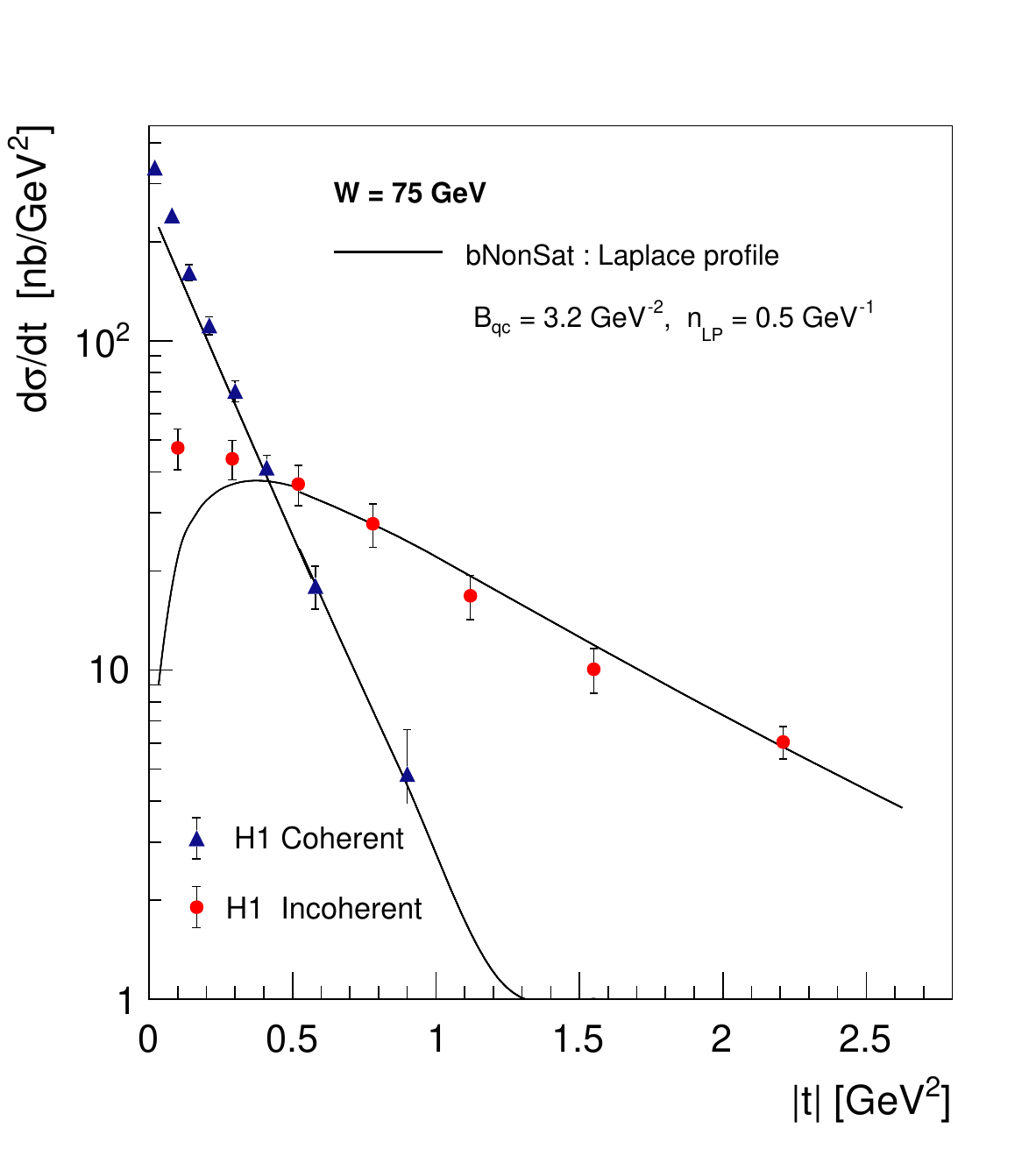}\hskip0.5cm
 		\includegraphics[width=0.5\linewidth]{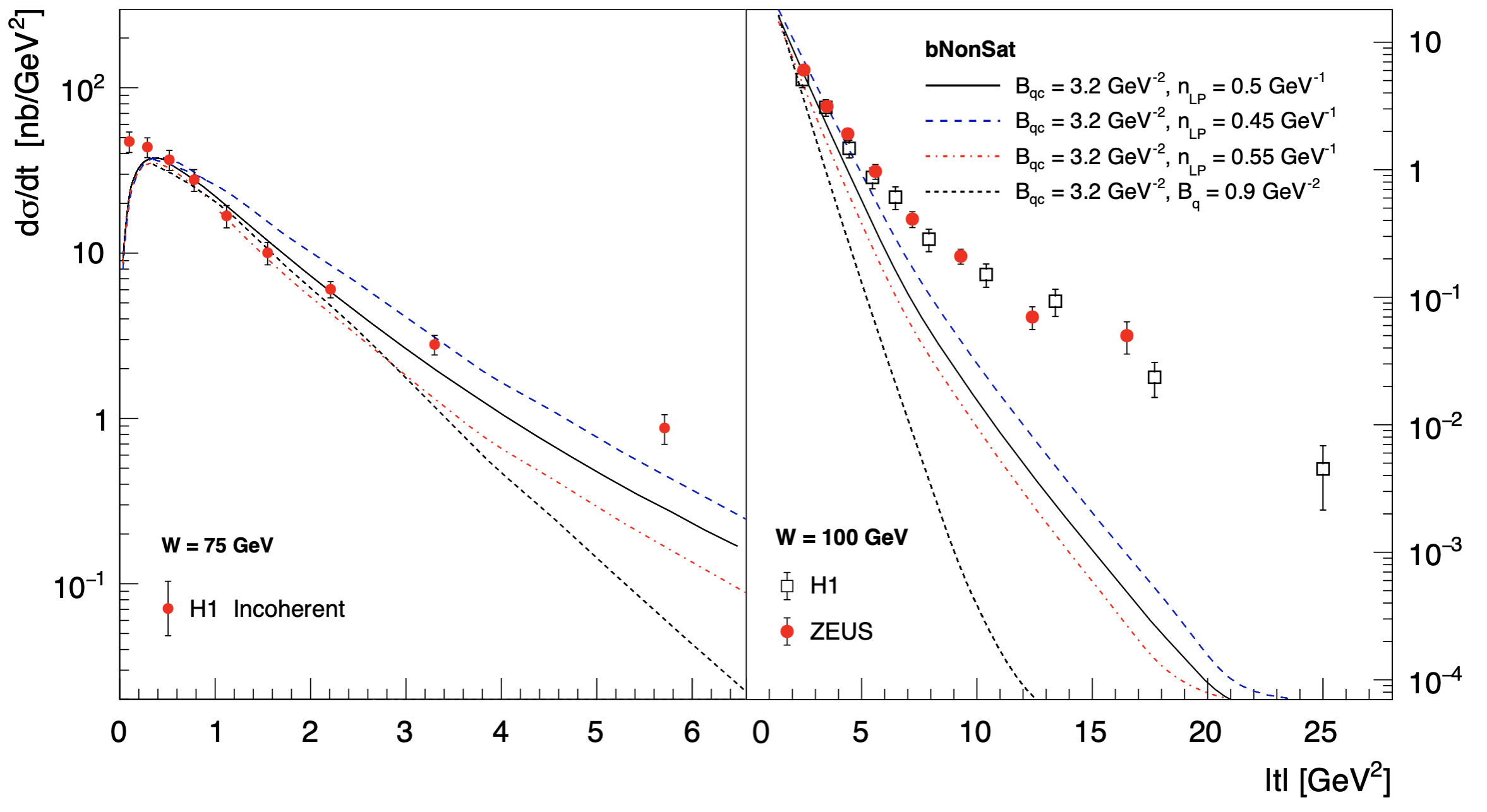}
 	\caption{The $|t|$ dependence of $J/\Psi$ photoproduction in the bNonSat hotspot model with Gaussian and Laplace  profile functions. In the left figure the model predictions are compared at low momentum transfer while on the right figure the incoherent cross section prediction are compared at large momentum region. The experimental data is taken from ZEUS \cite{Chekanov:2009ab} and H1 \cite{Aktas:2003zi}.   }
 	\label{laplace}
 \end{figure*}

In our model we have only considered Gaussian gluon distributions within the hotspots. If the probing dipole is large compare to the hotspot size we expect it to be able to resolve gluon-gluon correlations, which would modify the shape away from a Gaussian. In our model we have not introduced any correlations, but we may anticipate the effect of introducing these into the model by considering a non-Gaussian shape of the hotspots. This also helps to rule out that the need of further substructure in our description of the proton is an artifact from our choice of Gaussian shapes of the hotspots. As an alternative, we have chosen a Laplace profile for the hotspots. This gives a more peaked distribution in the centre as well as exponential tails, which when Fourier-transformed results in a power in $|t|$. When integrated over the $z$-direction, the  Laplace profile function for the hotspots becomes:
\begin{equation}
	T_{q}(\textbf{b}) = \frac{1}{4 \pi n_{LP}^3} b K_1\big[-\frac{b}{n_{LP}}\big]
\end{equation}
We found that the optimal values for the paramters are  $B_{qc} = 3.2$~ GeV$^{-2}$ and $n_{LP} = 0.5~$GeV$^{-1}$. This new choice of profile  explains coherent and incoherent data at low momentum transfer but fails at high momentum transfer as illustrated in Fig. \ref{laplace}. We find that while this profile function better describes the shape of the $|t|$ spectrum, and is able to describe the data for $|t|\lesssim 3$~GeV$^2$,  we still need to introduce substructure of the hotspots in order to describe the measurements at larger $|t|$.

\subsection*{Scaling properties of geometrical fluctuations}
\begin{figure}
	\centering
	\includegraphics[width=0.75\linewidth]{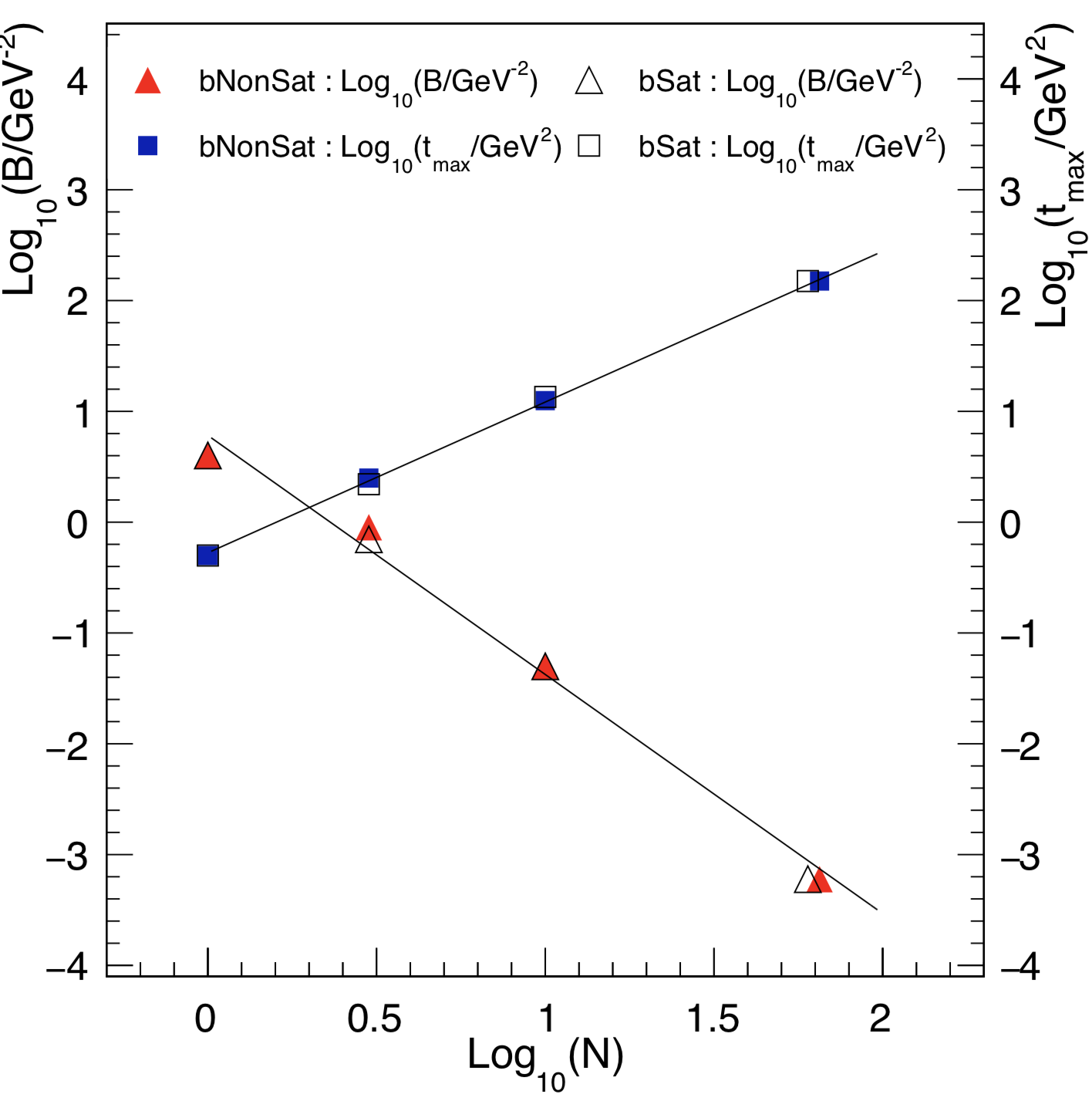}\hskip1.5cm
	\caption{Scaling behaviour of the spatial gluon density fluctuations  inside the proton of the parameters $B=\{B_G, B_q, B_{hs}, B_{hhs}\}$ and $N=\{1, N_q, N_{hs}, N_{hhs} \}$ . $|t|_{\rm max}$ is the upper value of $|t|$ to which the fluctuations at a scale $B$ contributes. For bSat: $|t|_{\rm max}=\{0.5, 2.2, 13.5,150\}~$GeV$^2$, and for bNonSat: $|t|_{\rm max}=\{0.5,  2.5, 12.5, 150\}~$GeV$^2$ where the last values are directly taken from the line.}
	\label{fig:scaling}
\end{figure}
Next we investigate the underlying structure of the subnucleon fluctuations. In Fig.~\ref{fig:scaling} we plot $\log(B)$ versus $\log(N)$ for our models, where each point represents a level of substructure, and the values for $B$ and $N$ are taken from table \ref{table}. For $|t|\lesssim0.5~$GeV$^2$ the coherent cross-section dominates, and the relevant scale is $B_G=4~$GeV$^{-2}$, while the larger hotspots contribute for $|t|\lesssim 2~$GeV$^2$ and as we increase the resolution we need smaller and smaller hotspots. As can be seen, these numbers fall on a line, which indicates that these parameters are highly correlated, and that there is a scaling in the structure of transverse gluon density fluctuations in the proton. 

We also show how
the upper value $|t|$ for which the fluctuations at a scale $B$ contributes. These are the $t$-values at which the incoherent spectrum changes slope. The last value is taken directly from the line in Fig.~\ref{fig:scaling}. We therefore expect the currently smallest hotspots in our model to stay relevant for $|t|$ up to $\sim150~$GeV$^2$ which is far higher than any currently conceivable measurement.

\section{Conclusions and Discussion}\label{conclusion}

The hotspot model for gluon fluctuations have proved to be an efficient model for describing the incoherent cross-section in $ep$ data, supplemented with large length-scale saturation scale fluctuations. However, as the hotspot model is non-perturbative, it is not reliable for $|t|\gtrsim 1$~GeV$^2$. As we probe the smaller scale fluctuations in incoherent diffraction at large $|t|$ we resolve more quantum fluctuations in the proton, and the $t$-spectrum is measured well into the perturbative regime. Further, in order to describe total exclusive diffractive cross-section, as well as rapidity and $W$ spectra which are integrated over $t$, one needs a description for incoherent diffraction at $|t|>1$~GeV$^2$. In lieu of a perturbative approach to this problem, we have added substructure to the proton by hand. This will tell us what we may expect from a perturbative approach, as well as enable us to calculate rapidity and $W$ spectra with the dipole model. 

We have shown that if we extend the hotspot model into a model which has hotspots within hotspots within hotspots, adding a structure of ten hotspots inside each of the original three and then further substructure with around sixty smaller hotspots inside each of the ten hotspots, we are able to describe the data for momentum transfer of at least $|t|<30~$GeV$^2$. We see that the gluon density fluctuation structure exhibit a scaling behaviour. If this scaling  persists, we expect our model with three levels of gluonic substructure in the proton to be relevant for $|t|\lesssim 150~$GeV$^2$. This scaling behaviour also reduces the degrees of freedom of the parameter space, as the parameters are highly correlated.

 In \cite{Mantysaari:2019jhh} M\"antysaari and Schenke use the Colour Glass Condensate (CGC) to investigate the deuteron incoherent cross section. The CGC contains gluon-gluon correlations, and they note that even if the CGC without subnucleon fluctuation gives a harder incoherent cross section compared to the bSat model, subnucleon fluctuations still make a significant contribution at larger $|t|$. 
In a recent paper \cite{Demirci:2022wuy}, the authors also use the hotspot model to describe the entire $t$-spectrum for $|t|<30$~GeV$^2$, using a Colour Glass Condensate assumption for colour fluctuations inside the hotspot. They very usefully take the non-relativistic and dilute limits in order to calculate the $t$-spectrum analytically and describe the full spectrum reasonably well. They find that the coherent description have a strong dependence on the probe as well as the target in $J/\psi$ production, while the incoherent spectrum is dominated by the target structure, which somewhat validates our approach. However, their resulting relative normalisations of coherent and incoherent cross-sections do not match the measurements. 
From this we conclude that even a more advanced model containing gluon-gluon correlations needs increased complexity as $|t|$ increases. Our findings in this paper systematically constrains which features such a model must have. 

We see that similarly to the large scale DGLAP limit, the phase-space density in the transverse plane becomes more dilute at large $|t|$. This may point towards a perturbative approach for extending the hotspot model to large $|t|$, similar to a DGLAP evolution of "hotspot splittings", taking the small $|t|$ hotspot model as a non-perturbative initial state.
 
Furthermore, we have introduced a modified hotspot thickness function that not only reproduces the small $|t|$ coherent spectrum, but also improve the small $|t|$ description in the incoherent cross section with one extra parameter which we interpret as a measure of gluon-gluon correlations. We therefore expect that this parameter will play a more significant role when we extend the investigation to $\rho$ and $\phi$ mesons, which contribute to the amplitude at larger dipoles than the $J/\psi$ mesons studied here.

Studies of the dependence of number of hotspots on momentum fraction $x$ suggest that either the number of hotspots, or their widths will increase with smaller $x$ (see \cite{Kumar:2022aly} for our recent detailed investigation of these effects). This would manifest in the $W_{\gamma p}$ dependence of the cross section. We see that our models describe well all available $W_{\gamma p}$ bins for all $t$. At large $|t|$ we would have a larger $x$ which may off-set the large $W_{\gamma p}$ behaviour in the proton substructure. A detailed study of the $W_{\gamma p}-t$ substructure of the proton would therefore be an intriguing endeavour both experimentally and phenomenologically. 

The future electron-ion collider, with its significantly increased luminosity compared to HERA, will be able to measure the incoherent $t$-spectrum at unsurpassed precision. Hopefully, the EIC will also be able to increase the reach in $t$, which would cast light on this compelling physics. This would for the first time enable a direct precision measurement of the scaling behaviour in the geometrical density fluctuations of gluons in the transverse plane.

\subsection*{Acknowledgements}
We thank T. Ulrich, T. Lappi, B. Schenke and H. M\"antysaari for fruitful discussions that has helped us to better understand our findings. The work of A.Kumar is supported by Department of Science \& Technology, India under DST/INSPIRES/03/2018/000344. We have used computing resources of our HEP-PH group at IIT Delhi and thank all the members of the group and the Physics Department of IIT Delhi.

\bibliographystyle{elsarticle-num}
\bibliography{bibliography}
    \end{document}